\documentclass[pra,aps,final,onecolumn,showpacs,nofootinbib]{revtex4-1}
\usepackage{epsfig}
\usepackage{latexsym}
\usepackage{xspace}
\usepackage{hyperref}
\usepackage[latin2]{inputenc}
\usepackage{indentfirst}
\usepackage{enumerate}
\usepackage{color}
\usepackage{colordvi}

\usepackage{amsmath}
\usepackage{amssymb}
\usepackage[english]{babel}
\usepackage{url}

\newcommand{\eq}[1]{\begin{align} #1 \end{align}}

\def\be{\begin{equation}}
\def\ee{\end{equation}}
\def\bea{\begin{eqnarray}}
\def\eea{\end{eqnarray}}
\def\l{\label}

\def\p{{\bf p}}

\def\r{{\bf r}}

\def\v{{\bf v}}

\def\d{\mbox{d}}

\def\siml{\;\hbox{\kern.1em \lower.7ex \hbox{$\sim$} \kern-1.12em
 \raise.5ex \hbox{$<$} \kern.1em}}
\def\simg{\;\hbox{\kern.1em \lower.7ex \hbox{$\sim$} \kern-1.12em
 \raise.5ex \hbox{$>$} \kern.1em}}

\begin{document}

\title{Velocity and absorption coefficient of sound waves in classical gases}

\author{A.G.~Magner} 
  \email{Email: magner@kinr.kiev.ua}
\affiliation{Institute for Nuclear Research, National Academy of Sciences
of Ukraine, Kiev, Ukraine 03680}

\author{M.I.~Gorenstein} 
\affiliation{
Bogolyubov Institute for Theoretical Physics, National Academy of Sciences
of Ukraine,, Kiev, Ukraine 03143}

\author{U.V.~Grygoriev}
\affiliation{
Institute for Nuclear Research, National Academy of Sciences
of Ukraine, Kiev, Ukraine 03680}

\begin{abstract}
Velocity and absorption coefficient
of the plane sound waves in classical gases
are obtained by solving the Boltzmann kinetic equation.
This is done within the
linear response theory as a reaction of the single-particle
distribution function
to  a periodic external field.
The nonperturbative
dispersion equation is derived in the relaxation time
approximation and solved numerically.
The obtained theoretical results demonstrate
an universal dependence
of the sound velocity and scaled absorption coefficient on
variable $\omega\tau$, where
$\omega$ is the sound frequency and $\tau^{-1}$ is
the particle collision frequency.
In the region of $\omega\tau\sim 1$ a
transition
from
the frequent- to rare-collision regimes takes place.
The sound velocity increases
sharply,
and the scaled
absorption coefficient
has a maximum -- both theoretical findings are
in agreement with the data.

\vspace{0.2cm}

\vspace{0.2cm}
\end{abstract}

\maketitle

\section{Introduction: hydrodynamics and kinetics}

Sound waves in classical gases
were studied intensively within the hydrodynamical approach
(see, e.g., Ref.~\cite{LLv6}).
The sound velocity $c_0$ in this approach is equal to
\eq{\label{speed-sound}
c_0~=~\left(\frac{c_p\,k^{}_BT}{c_v\, m}\right)^{1/2}~,
}
where $k^{}_B$ is the Boltzmann constant,
$T$ is the system temperature,
$m$ the particle mass, $c_p$ and $c_v$ are the specific
heat capacity  at constant pressure and
constant volume, respectively.
The sound velocity (\ref{speed-sound}) appears to be independent of
the sound wave frequency $\omega$ and approximately equals to the thermal
particle velocity.
For absorbed plane sound waves (APSW)
the wave amplitude  decreases  as
$\exp(-\gamma z)$ after propagating the distance $z$.
The absorption coefficient
$\gamma$ is usually evaluated
from the Stokes relation \cite{LLv6},
\eq{\l{Stokes}
\gamma=\frac{\omega^2}{2 m n c^3_0}
\left[\frac43 \eta + \zeta +
\left(\frac{1}{c_v}-\frac{1}{c_p}\right)~\kappa \right]\;,
}
where $n$ is the particle number density,  $\eta$
the shear viscosity,
$\zeta$
the bulk viscosity, and $\kappa$ is the thermal conductivity.

Within a hydrodynamic approach, the kinetic coefficients
are phenomenological constants.
For calculations of the kinetic coefficients in Eq.~(\ref{Stokes}),
one needs the kinetic theory.
The global equilibrium of a classical gas is described
then by the Maxwell distribution function of particle's momentum $\p$ with
$p\equiv |\p|$:
\be\l{M}
f^{}_{\rm GE}(\p)~=~
\frac{n}{(2\pi m k^{}_BT)^{3/2}}\,
\exp\left(-\;\frac{p^2}{2m k^{}_BT}\right)\;.
\ee
In this equation,
the particle number density $n$ and temperature $T$
are independent of the spacial coordinates
$\r$ and time $t$.  The equilibrium in a classical gas is achieved
by successive two-body collisions with the elastic cross-section $\sigma$
equal
to $\pi d^2$ for  hard-sphere particles with a diameter $d$.

The average value of thermal  velocity $v$ can be calculated from
Eq.~(\ref{M})  as ($\p \equiv m\v$ and $v\equiv |\v|$):
\eq{\label{v-av}
\overline{v}~=~\frac{\int p^2dp~(p/m)~f_{\rm GE}(p)}{\int p^2dp~f_{\rm GE}(p)}~
=~\left(\frac{8 k^{}_BT}{\pi\, m}\right)^{1/2}~.
}
The particle mean-free path can be found analytically as a
function of $\sigma$ and $n$ \cite{To66,Fe72}:
\eq{\label{l-1}
l~=~\left(\sqrt{2}\sigma\,n\right)^{-1}~.
}
From Eqs.~(\ref{v-av}) and (\ref{l-1}) one finds the collision frequency
as
\eq{\label{tau1}
\tau^{-1} ~\equiv~\overline{v}/l~=~4\sigma\,n~\sqrt{\frac{k^{}_BT}{\pi\,m}}~.
}
In terms of the above quantities the shear viscosity $\eta$ can be calculated as
\eq{\label{eta-1}
\eta~=~\frac{5\pi}{32}~n\,m\,l\,\overline{v}~
=~\frac{5\sqrt{\pi}}{16}~\frac{\sqrt{m k^{}_BT}}{\sigma}~.
}
The numerical coefficient in Eq.~(\ref{eta-1}) was found by
Chapman and Enskog (see, e.g., Ref.~\cite{chapman}).
The thermal conductivity can be then found as
\eq{\label{kappa}
\kappa~=~\frac{15}{4}\eta~,
}
and the bulk viscosity for non-relativistic mono-atomic
gases equals to zero, $\zeta=0$
(see, e.g., \cite{LPv10}).

The molecular kinetic scheme based on the above
equations is
self-consistent for  dilute gases when
the mean free path $l$ is much larger than the size of
particles, $l\gg d$. For this case,
the excluded volume effects due to the particle hard-core
repulsion appear to be negligible,
i.e., the gas pressure, $P=n k^{}_BT$, and specific heat capacity,
$c_v=3/2$ and $c_p=5/2$, are equal to their ideal gas values with
high accuracy \cite{LLv5}. In what follows 
the thermodynamical relations
(\ref{v-av}-\ref{tau1})
are assumed to be valid,
i.e., our consideration is restricted to the case of dilute classical gases.
Equations ~(\ref{eta-1}) and (\ref{kappa}) are the leading terms of the
perturbation expansion over small
Knudsen parameter, ${\mathcal K}\equiv \omega\tau \ll 1$.
This corresponds to the so-called frequent
collision regime (FCR).

\begin{figure}
\begin{center}
  ~~~~\includegraphics[width=0.77\textwidth,clip]{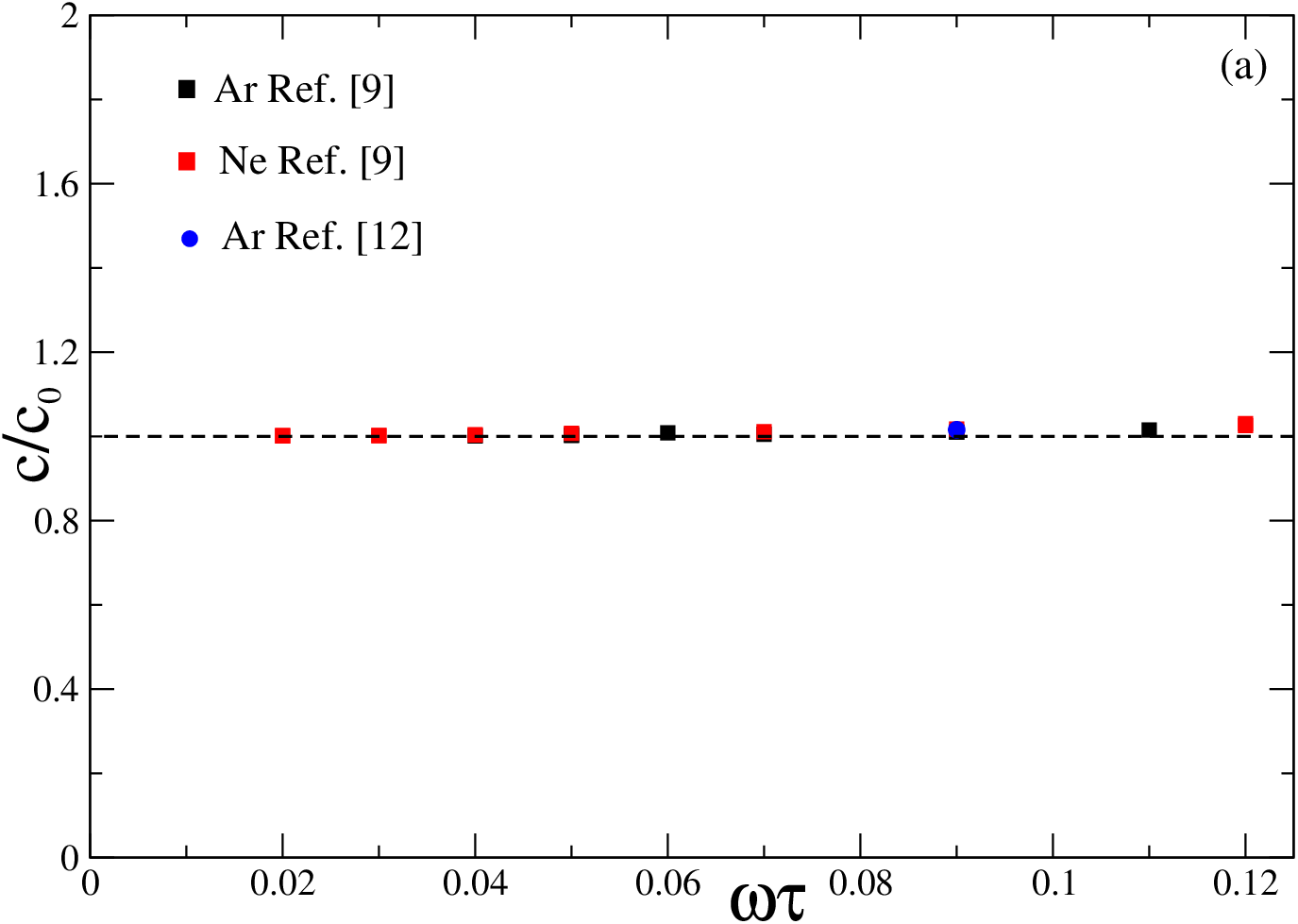}
 
\includegraphics[width=0.8\textwidth,clip]{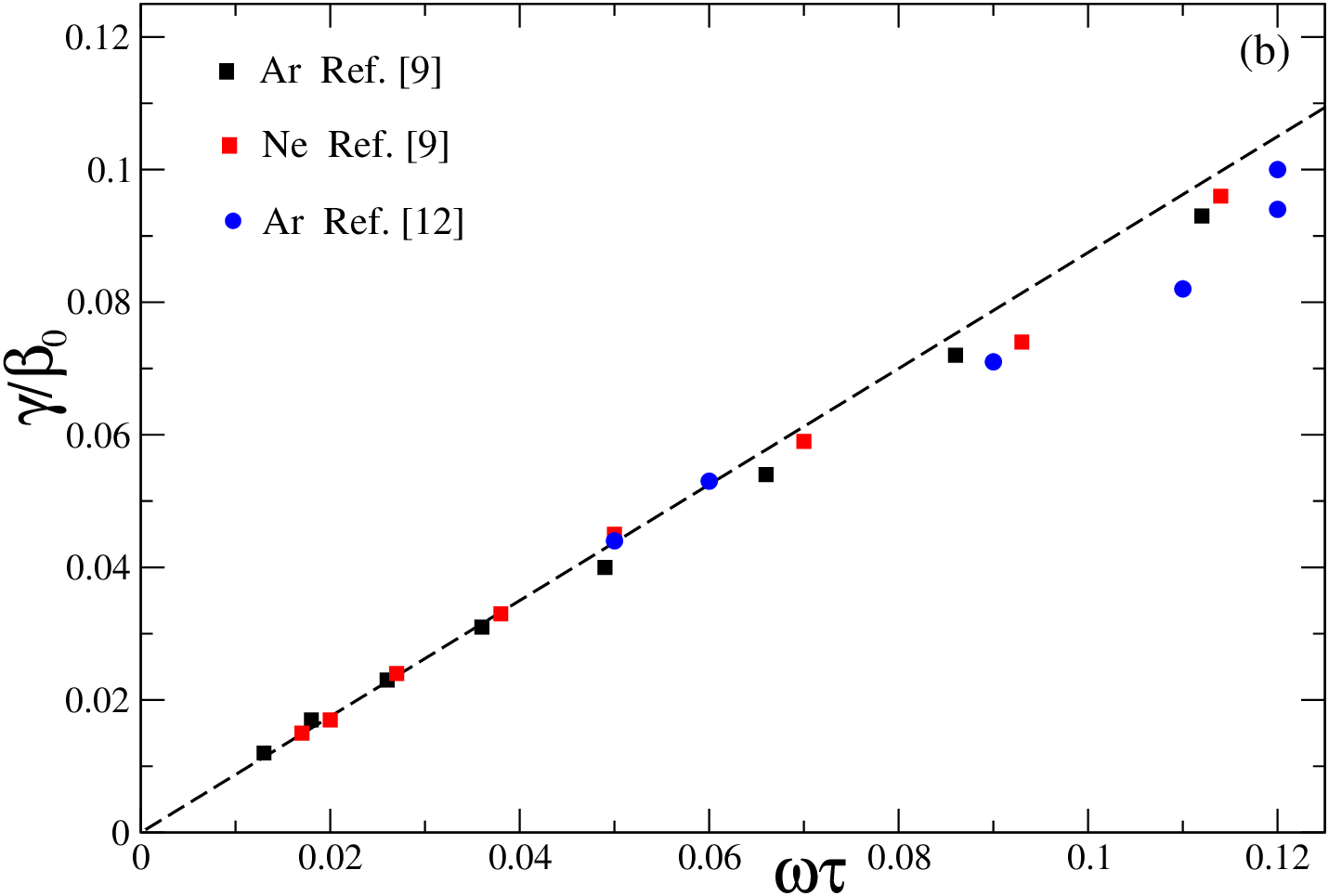}
\end{center}
\caption{
  The speed of sound $c/c_0$ (a) and
the scaled absorption coefficient
$\gamma/\beta^{}_0$ (b)
as functions of $\omega\tau$ at low frequencies.
 The experimental data
are taken from Refs.~\cite{Gr56,Me57}.
Dashed lines present Eq.~(\ref{speed-sound}) in (a)
and the Stokes formula (\ref{stokes-1}) in (b).
}
\label{fig1}
\end{figure}

Using the above equations,
one finds for  the FCR scaled absorption coefficient from
the Stokes formula (\ref{Stokes}):
\eq{\label{stokes-1}
\frac{\gamma}{\beta_0}~=~\frac{7}{8}\,\omega\tau~,
}
where $\beta_0\equiv \omega/c_0$ is the wave number. As shown
in Fig.~\ref{fig1},
both $c_0$ and $\gamma/\beta_0$
given by
Eqs.~(\ref{speed-sound}) and
(\ref{stokes-1}), respectively, are supported by the data
at small $\omega\tau$. 
%
In most practical cases, the
inequality $l \ll \lambda $, where
$\lambda \cong 2\pi c_0/\omega$ is the sound wavelength, is satisfied,
and, thus, the FCR is valid.
For example, for gases at normal conditions one gets
$\omega\tau\sim 10^{-8}-10^{-5} $ for the audible frequency region.
Calculating $\gamma$ from Eq.~(\ref{stokes-1}){\bf ,} one finds
$\gamma^{-1} \sim l(\omega\tau)^{-2}\sim 10^5-10^8$~cm
($c^{}_0\sim \overline{v}$), i.e.,
the audible APSW  propagate $1-10^3$ kilometers before
its amplitude decreases by the factor of $e^{-1}$.
The absorption of these waves is indeed rather weak.
Note also that in the FCR all kinetic coefficients
depend only on the equilibrium gas quantities. 
For example, the shear viscosity (\ref{eta-1}) and
thermal conductivity (\ref{kappa}) are independent of the sound frequency.

Equations ~(\ref{speed-sound}) and 
(\ref{stokes-1}) 
are, however,  in a contradiction with
the existing data at $\omega\tau \simg 1$; see critical comments, e.g., in
Refs.~\cite{Fe72,Ce69,Ce75}.
This is a transition region from the FCR to the rare-collision regime (RCR).
The RCR corresponds to large values of the Knudsen
parameter,
${\mathcal K}\equiv \omega\tau \gg 1$.
The conditions of the RCR emerge at a small
particle number density (or small pressure), where  $l$ increases as $n^{-1}$
by Eq.~(\ref{l-1}),
and for large sound-wave frequency $\omega$.
The basic experiments for sound waves in classical gases
at large
values of $\omega\tau$ (the so-called ultrasonic waves) were done by
Greenspan \cite{Gr56,Gr65book,schotter74} and Meyer-Sessler \cite{Me57}.
Many works were devoted to
a comparison of the theoretical and experimental results
for the sound velocity and absorption. One of the direction
on the early study of this problem was related to the accounting for
high ${\bf p}$-moments of Boltzmann kinetic equation (BKE) within the
method of  moments,
called the Bartnett \cite{WaCh48} and super-Bartnett \cite{WaCh52}
models to obtain solutions for $\omega\tau<1$,
see also Refs.~\cite{Ce69,Ce75,Wo80,Le89,Ko67}.
An alternative kinetic approach
is based on the  well known Bhatnagar-Gross-Krook 
model \cite{BGK54} and its generalisation by Gross and Jackson \cite{GJ59},
see also Refs.~\cite{Fe72,Ce75}.
The main idea was to use the linearized BKE with a relaxation $\tau$
approximation for the integral collision term and express exactly
the solution of this equation in terms of local dynamical variations
of the particle number density,
$\delta n({\bf r},t)$, the velocity field ${\bf u}$ and temperature variations
$\delta T({\bf r},t)$. Using the
conditions to restore the conservation
relationships (violated, generally speaking, by the $\tau$ approximation, see
also Refs.~\cite{He93,Ta10,Ma14}),
one can then
approximately obtain the dispersion equation
for the sound velocity \cite{Ta10}.
They suggested also to apply the boundary conditions
on the walls of a conductor pipe for particle gas motion. Such solutions
were found
\cite{Fe72,Ce75,Si65,Bu66,Th79,Lo79,Ch81,Ga06,Sh08} for the
simplest case of the diffused boundary conditions on a plane surface of
the semi-infinite gas system. Usually, these boundary conditions
for the distribution function are considered by assuming mirror or
diffuse reflections
of particles from the boundary \cite{Be61,Iv81,Ma87}. The linearisation
of the BKE for calculations of the sound velocity and absorption
in the solutions
in terms of the plane wave can be
grounded for the case of a small influence of boundary conditions of
the sound propagation for a size of the conductor pipe
much larger than the wavelength; see, e.g., Ref.~\cite{Ha01}
for such numerical BKE solutions for the sound velocity
and absorption without using the boundary condistions.
As shown in Ref.~\cite{St87} for the kinetic equation with
mirror boundary conditions
taking them at the plane surfaces of a slab and in Ref.~\cite{Ko94}
for the monopole vibrations with diffuse boundary conditions
at the spherical surface, the sound solutions are, in fact, rather different
from a plane wave because they depend very much on the type of the
boundary conditions and
choice of the boundary geometry itself.

Both the FCR and RCR for the sound velocity and absorption analytically in
terms of a simple plane wave without using different boundary conditions
and, therefore, independent of their specific properties
have been studied within the linearized Boltzmann kinetic equation (BKE)
in our recent paper \cite{MGGpre17}.
The approximate expansions for
small and large values of   ${\mathcal K}$ were obtained analytically
 by using
the $\tau$ approximation
for the integral collision term with a constant $\tau$, that is
independent of the
particle velocity.
The dispersion equation for the sound velocity and absorption
was derived approximately analytically within the linear response
theory  \cite{MGGpre17a}
following the ideas of the BGK model \cite{BGK54}.
By solving this dispersion equation numerically for 
any value of  ${\mathcal K}$, one obtained the sound velocity and absorption
for a transition from the FCR to the RCR.
 In the present paper, our theoretical approach 
is worked out and
compared carefully with the available data.

\section{Boltzmann kinetic approach}

We consider
the BKE
for the 
single-particle distribution function
$f(\r,\p,t)~$ of the coordinate $\r$,
momentum $\p$ and time $t$
(see e.g., Ref.~\cite{LPv10}), 
with the external potential
$V_{\rm ext}(z,t)$ (more details can be found in Ref.~\cite{MGGpre17a}),
\be\label{Vext}
V_{\rm ext}(z,t)=
\exp\left(- ~ i \, \omega \,t
\right)
\int_{-\infty}^\infty \frac{\d k}{2 \pi}~ V_k \;
\exp\left(i k z\right)\;,
\ee
where $V_k$ is the Fourier amplitudes\footnote{As usual, the complex number representation is used for
convenience, but only the real parts of $f$ and $V_{\rm ext}$ will be taken as
physical quantities.}.
The linearized BKE takes  the form:
\be\l{Boltzlin}
\frac{\partial \delta f}{\partial t}+
\frac{p_z}{m} \frac{\partial \delta f}{\partial z}
-\delta St[f] =
\frac{\partial f^{}_{\rm GE}}{\partial p_z}\;
\frac{\partial V_{\rm ext}}{\partial z}\;,
\ee
where
$\delta f(\r,\p,t)\equiv f(\r,\p,t)-f^{}_{\rm GE}(p)$,
and the
collision term $\delta St[f]$
is taken in the standard Boltzmann
form (see, e.g., Ref.~\cite{Fe72}).

A small periodic external field (\ref{Vext})
induces the corresponding 
deviations
\eq{
\delta f(z,\p,t)~ =~\exp(-i \omega t) \int_{-\infty}^{\infty}\frac{dk}{2\pi}\,
f_k~\exp\left(i\, kz\right) 
\l{dfk}
}
with small Fourier amplitudes
$
f_k \propto V_k$
(see, e.g., \cite{Ma14,Fo75,Zu02}).
Equation (\ref{Boltzlin}) is assumed to
be valid at $|\delta f|/f^{}_{\rm GE}\ll 1$ 
in the APSW 
form 
with the frequency $\omega$
propagated  along the $z$ axis.
For the integral collision term $\delta St$ we use the 
$\tau$ relaxation-time approximation 
in the form \cite{MGGpre17a}:
\bea \l{tauapprox}
\delta St[f]~\cong~
-\frac{1}{\tau}~\left(\delta f-\delta f_{\rm LE}\right)~
\equiv~-~\frac{1}{\tau}\,\delta\varphi~,
\eea
where $\tau^{-1}$ is constant given by Eq.~(\ref{tau1}), and
the local equilibrium part of $\delta f$ is related to
the well known Maxwellian function,
\eq{\label{le}
\delta f^{}_{\rm LE}
= f^{}_{\rm GE}\,\left[ \frac{\delta n}{n}+p_z \frac{\delta u_z}{k^{}_BT}+
\left(\frac{p^2}{2m k^{}_BT}-\frac{3}{2}\right) \frac{\delta T}{T} \right]~.
}
In Eq.~(\ref{tauapprox}),
$\delta\varphi
\equiv \delta f-\delta f_{\rm LE}$ appears as the
additional component responsible for
a sound absorption in a gas
through the
integral collision term $\delta St[f]$ (\ref{tauapprox}).
Note that one has $St[f_{\rm LE}]=0$ at the
local equilibrium distribution function
$f_{\rm LE}$ for any parameters $\delta n$, $\delta T$, and $\delta u_z$
\cite{Fe72,Si71}. Thus, just the $\delta \varphi$ term is responsible
for all dissipative effects in a
gas.
In Eq.~(\ref{le}), the variations
$\delta n$, $\delta T$, and $\delta u_z$
are small deviations
of the particle number density, temperature,
and collective velocity, $|\delta n|/n \ll 1$, $|\delta T|/T\ll 1$,
and $|\delta u_z|/\overline{v} \ll 1$, from their GE values $n$, $T$,
and $u_z=0$.
The conservation of particle number,  momentum, and energy impose the following
requirements
\cite{Ta10,MGGpre17a}:
\eq{
& \int d\p\,\delta \varphi~=~0~,~~~~~
\int \d\p\, p_z~\delta \varphi~=~0~,\label{nu-cons}\\
& \int d\p\,p^2\,\delta \varphi~=~0~. \label{en-cons}
}
In what follows, for simplicity, we put $\delta T=0$ in Eq.~(\ref{le}),
i.e., the effects
of thermal conductivity will be neglected.
For  constant temperature $T$, only Eq.~(\ref{nu-cons})
should be considered. Equation  (\ref{en-cons})
for the energy conservation 
is then identically satisfied.

The solution of the linearized BKE (\ref{Boltzlin})
is found from Eq.~(\ref{dfk}) by calculating the $k$-integral
by the residue method in the following APSW form:
\eq{
\delta f(z,\p,t)
\propto  \exp\left(-i\omega t +ik_0z\right)~,
\l{dfk-1}
}
where the poles in the complex $k$ plane are connected with the speed
of sound $c$ and absorption coefficient $\gamma$, e.g., as
\eq{\label{k0}
k^{}_0~=~\frac{\omega}{c} + i\gamma~.
}
The position of poles $k^{}_0$
are obtained from the following dispersion relation \cite{MGGpre17a},
\eq{\l{D}
\mbox{D}(w,\mathcal{K})
\equiv
\left[
\frac{i\,w}{\xi\,\mathcal{K}}\;\left(1+Q_1\right)
-1\right]
\left(3 i\,\frac{w}{\mathcal{K}}\, \xi \,
Q_1-1\right)
+\frac{8}{\pi}\;\left(\frac{w\, Q_1}{\mathcal{K}}\right)^2
~=~0~,
}
where $w
\equiv \omega/k^{}_0 v^{}_T=w_r+i w_i$
with $v^{}_T\equiv (2 k^{}_BT/m)^{1/2}$, and
$Q_1(\xi)\equiv (\xi/2){\rm ln}[(\xi+1)/(\xi-1)]-1$
with $\xi \equiv  w(1+i/\mathcal{K})$.  In the analytical derivation
  of this dispersion equation we used approximately $p\approx p^{}_T=m v^{}_T$
by  the properties of the Maxwellian distribution (\ref{M})
calculating angle integrals over the momentum ${\bf p}$, that simplifies
this equation, in contrast to the derivations in Ref.~\cite{Ta10}.
The absolute values of the sound-wave number $\beta=\omega/w_r$
and the scaled  absorption coefficient $\gamma/\beta$
are given by
\be\l{krdef}
\beta
~=~\frac{\omega}{v_T}~
\frac{|w_r|}{w_r^2+w_i^2}\;,~~~~~~
\frac{\gamma}{\beta}~=~\Big|\frac{w_i}{w_r}\Big|~.
\ee
Thus, one obtains the wave number $\beta>0$
and the absorption
coefficient $\gamma>0$ for sound waves spreading in the
positive $z$-axis direction  for $z>0$. Similarly, one finds the contributions
of other poles \cite{MGGpre17a}.

Taking the asymptotic expansion
of $\mbox{D}(w,\mathcal{K})$
[Eq.~(\ref{D})] in a power series
over $\mathcal{K}$,
within the FCR
where $\mathcal{K} \ll 1$, for the isothermal sound velocity $c$ in
units of the
adiabatic sound velocity $c^{}_0$ [Eq.~(\ref{speed-sound})], one finds
\bea \label{c-FCR}
\frac{c}{c^{}_0}
&\cong& \frac{4}{\sqrt{15\pi}} + a_2 (\omega\tau)^2
+O\left[(\omega\tau)^4\right]~,\\
\frac{\gamma}{\beta^{}_0} &\cong&
\frac{(21\pi-40)\sqrt{15\pi}}{160}~\omega\tau
+O\left[
(\omega\tau)^3\right]~,\l{FCR}
\eea
where $a^{}_2 \cong
0.67~$ ($\beta^{}_0=\omega/c^{}_0$).
In the RCR, $\mathcal{K} \gg 1$,
one
obtains from the asymptotic expansion of  Eq.~(\ref{D})
over $1/\omega\tau$
\bea\l{c-RCR}
\frac{c}{c^{}_0} &\cong& \left[1~-\frac{1}{(\omega\tau)^2}\right]
~\sqrt{\frac{6}{5}}
+
O\left[(\omega\tau)^{-4}\right]~,\\
\frac{\gamma}{\beta^{}_0}&\cong&
\sqrt{\frac{5}{6}}~ \frac{1}{\omega\tau}+
O\left[(\omega\tau)^{-4}\right]~.\label{RCR}
\eea
%

\begin{figure}
\begin{center}
\includegraphics[width=0.8\textwidth,clip]{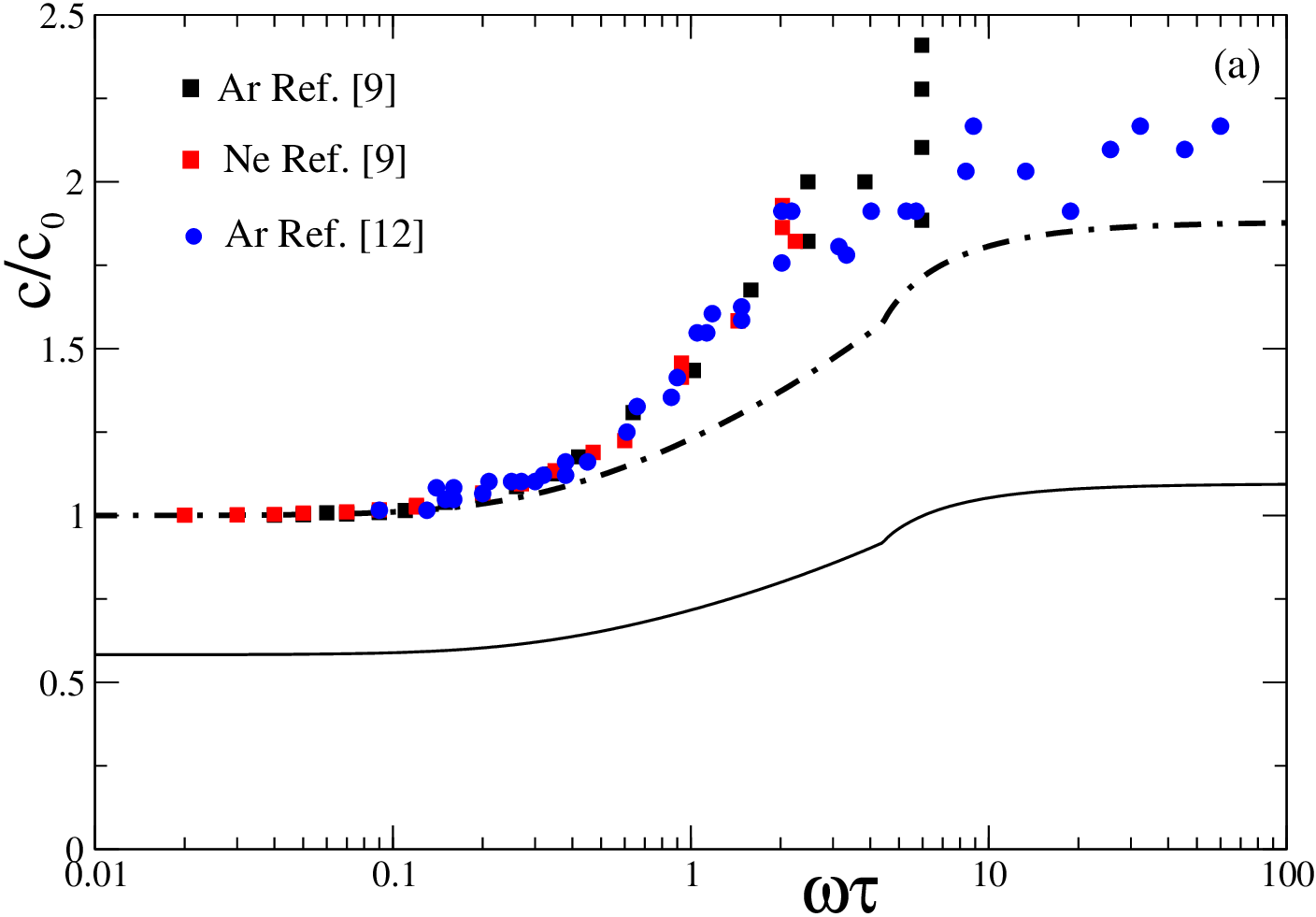}
\includegraphics[width=0.8\textwidth,clip]{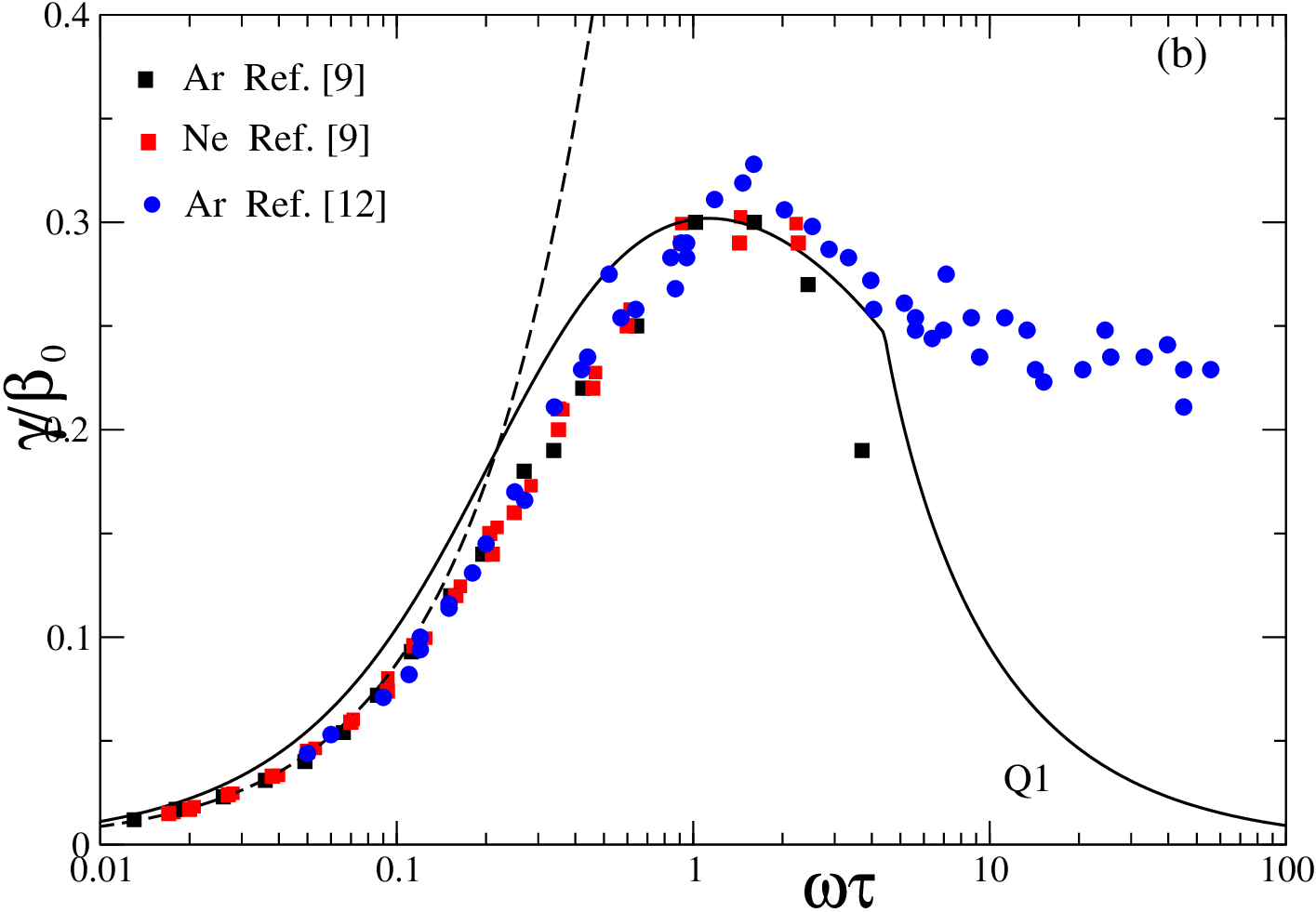}
\end{center}
\caption{
Sound velocity $c$ (a)
in the units of $c_0$ (\ref{speed-sound}) and
the scaled absorption coefficients $\gamma/\beta^{}_0$ (b)
as functions of $\omega\tau$.
Solid black lines show the
nonperturbative  solutions to the dispersion equation (\ref{D}). 
Dash-dotted line in (a) 
is the relative sound velocity $(c/c^{}_0)(15\pi)^{1/2}/4$.
Dashed line in (b) presents
the Stokes equation (\ref{stokes-1}) calculated in the  FCR.
}
\label{fig2}
\end{figure}

\section{Comparison with data}
In this section we present a comparison of the obtained
theoretical results
with the available data  \cite{Gr56,Me57}
on the speed and
absorption coefficient of  
APSWs. These data
were originally  presented in terms of the dimensionless variable
\cite{Gr56,Gr65book,Me57}
\eq{\label{r}
r~=~\frac{2f_c}{5\pi\, f}~\equiv~\frac{4f_c}{5\omega}~,
}
where $f_c=\overline{v}/l$, which
coincides with $\tau^{-1}$
given by Eq.~(\ref{tau1}).
We prefer to use the quantity $\omega\tau\equiv 0.8 r^{-1}$.
Thus, in contrast with the
original presentation, the frequency $\omega$ in our figures
is increasing from left to right
along the abscissa axis.

The fundamental experimental observation consists
in the following.
 All mono-atomic gases
with different
values of a mass $m$ and
diameter $d$ (cross-section $\sigma=\pi d^2$),
and of a
density $n$ and temperature $T$, i.e., different
equilibrium states of the gas,  lead to the same universal behavior
of the speed of sound and scaled absorption coefficient
presented in terms of the dimensionless quantity $\omega\tau$.
The data at different
$m$, $\sigma$, $T$, and $n$ values  can be shown
at the same figure,
but with $\tau=\tau(m,\sigma,n,T)$ calculated
by Eq.~(\ref{tau1}).
One should only remember that the same value of
$\omega\tau$ may correspond to different $\omega$
and $\tau=\tau(m,\sigma,n,T)$ values. Our  
description is in complete agreement
with this experimentally observed scaling.
Note that in
Ref.~\cite{Gr56} the pressure $P$ was changed from 1~atm to
$10^{-2}-10^{-3}$~atm. In addition to Ar and Ne atoms presented in
Figs.~\ref{fig1} and \ref{fig2},
the measurements were also done for He, Kr, and Xe.

Figure \ref{fig2}
shows
the sound velocity $c/c_0$ (a)
and scaled absorption coefficient
$\gamma/\beta_0$ (b)
as functions of the Knudsen parameter $\omega\tau$.
The results presented by solid lines
are obtained numerically by
solving the dispersion equation (\ref{D}).
 In the both limits $\omega\tau \ll 1$ and
$\omega\tau \gg 1$, our numerical results converge
\cite{MGGpre17a} to the asymptotic results
of the FCR and RCR, respectively. These limiting behaviors
correspond
to  Eqs.~(\ref{FCR}) and
(\ref{RCR}) at leading (quadratic) orders.

As seen from Fig.~\ref{fig2} (a), $c/c_0$ increases
in a transition region from the FCR to the RCR
where $\omega\tau \sim 1$.
This theoretical result is in a qualitative
agreement with the data.
Figure~\ref{fig2} (a)
shows however the numerical discrepancies for the absolute values of
$c/c^{}_0$. Their main reason is a well-known difference between
isothermal and adiabatic limits of the sound velocity \cite{LLv6}.
Comparison with the
experimental results 
is improved 
  for a relative sound dispersion  \cite{Ko67},
  $(c/c^{}_0)(15\pi)^{1/2}/4$, 
  i.e., for
the $\omega\tau$-dependent deflection of
the sound velocity $c$
from its FCR  limit, see the dash-dotted
line in Fig.~2 (a).

Figure \ref{fig2} (b) shows
the scaled absorption coefficient
$\gamma/\beta_0$  [Eq.\ (\ref{krdef})].
The
absorption coefficient $\gamma/\beta_0$  as a
function of the Knudsen parameter
demonstrates  a maximum at $\omega\tau \sim 1$
in the transition from the FCR to the RCR.
This behavior
is also in qualitative
agreement [see Fig.~\ref{fig2} (b)] with
the experimental data \cite{Gr56,Me57}. For even larger
$\omega\tau$, one finds a
significant difference from the results of Ref.~\cite{Me57}.
The ``kink'' point in the dependence of the scaled absorption coefficient
$\gamma/\beta$ and of the sound velocity $w_r$ on the Knudsen
 parameter $\mathcal{K}$ is found numerically on right of the maximum at
 $\omega\tau\approx 4.47$, where their derivatives  
with respect to  
$\omega\tau$  are sharply changed.
This is  
obtained in the numerical calculations which 
are carefully 
checked within two different numerical schemes.
It is interesting to note that the branch point (b.p.) of the
Legendre function $Q_1(\xi)$ of second kind (zero of its
logaritmic presentation) in the dispersion equation
(\ref{D}) is determined by $\xi=1$, that means for the sound velocity
$w_{\rm b.p.}=\mathcal{K}/(\mathcal{K}+i)$.
This corresponds to
the scaled absorption $\gamma/\beta$ at the branch point
$(\gamma/\beta)_{\rm b.p.}=1/\mathcal{K}$.
However, we did not find analytically
the Knudsen parameter $\mathcal{K}$ at the branch point.  
There are two 
length scales in the problem: the mean free path of particles 
in a gas, $l=v^{}_T\tau$, 
and the sound 
wavelength, $\lambda=w_rv_T 2\pi/\omega$.   The ``kink'' corresponds to
the $\omega\tau$ point where these two different scales
become approximately equal,
$l\approx \lambda$.
It takes place in the nonperturbative region of
$\omega\tau$ values. Nevertheless, the branch point $w_{\rm b.p.}(\mathcal{K})$
as function of $\mathcal{K}$ coincides
with the asymptotic RCR solution $w^{}_{RCR}=\mathcal{K}/(\mathcal{K}+i)$
found in Ref.~\cite{MGGpre17a} for large $\mathcal{K}$  and,
thus, one obtains approximately
$\gamma/\beta=1/\mathcal{K}$ starting from this ``kink'' point.
Such a presence of the ``kink'' point resembles a situation similar  to
phase transitions
in statistical mechanics. An origin of the ``kink'' remains the open problem
that deserves further studies.

Note that in the RCR, $\omega\tau \gg 1$, one finds
$\gamma^{-1}\sim l$ from Eq.~(\ref{RCR}),
i.e., the propagating
length of the plane sound waves is of the order of a mean-free path
in a gas. The quantity $l$ [Eq.~(\ref{l-1})]
remains rather small even for dilute gases. In a gas at the normal conditions,
the mean-free path is estimated as $l\sim 10^{-5}$~cm.
Even at much small pressures, e.g., $P \sim 10^{-3}$~atm, the
propagating length $\gamma^{-1}\sim l\sim 10^{-2}$~cm remains
in fact rather small at $\omega\tau \gg 1$ .
On the other hand, the Stokes formula
(\ref{Stokes}) was obtained from the transport equation for
the entropy density \cite{LLv6} under the assumption of
a weak absorption,
$\gamma \lambda/(2\pi) \ll 1$.
In the RCR this estimate is still valid
as $\lambda \ll 2\pi l$. Therefore, one may expect a validity of
Eq.~(\ref{Stokes}) in the RCR, too. This 
requires, however, a strong modification of the kinetic
coefficients which become also dependent on the sound frequency
$\omega$ (see Ref.~\cite{MGGpre17}).

\section{Summary}

The kinetic approach based on the linear response theory
for the BKE is developed
to calculate the velocity and absorption coefficient for the
plane sound waves.
Our solution
is based on the relaxation time approximation for the Boltzmann
collision integral term in the  classical dilute gases.

Nonperturbative numerical solutions
are found for the sound velocity
and absorption coefficient as functions of the Knudsen parameter
$\omega\tau$. It agrees with the asymptotic expansions in both FCR and RCR
approximations. Our results are  in
agreement with experimentally observed scaling
which means a dependence of both sound wave quantities -- velocity
 $c/c_0$ and absorption coefficient $\gamma/\beta^{}_0$ --
from a single dimensional parameter $\omega\tau$.
Qualitative changes of the sound velocity and scaled
absorption coefficient  in the transition region $\omega\tau\sim 1$
are observed:
The sound velocity $c/c_0$ strongly increases
while the absorption coefficient $\gamma/\beta^{}_0$
has a maximum
at $\omega\tau \approx 1$.
Both these theoretical results are in
agreement with the data.

Our theoretical description is  not complete.
The presented
BKE calculations
should be extended to account for
the thermal
conductivity effects.
In the
RCR, the experimental values of $\gamma/\beta_0$ seem to be essentially
larger than our
estimate $\sim (\omega\tau)^{-1}$ (\ref{RCR}) at $\omega\tau\gg 1$.
This  can
be a signal of a presence of additional physical mechanisms for the
sound-wave suppression, that were not included in the
present formulation.

\vspace{0.2cm}
{\centerline{\bf Acknowledgements}}
\vspace{0.2cm}

We thank S.N.\ Reznik,
P.\ Ring, and  A.I.\ Sanzhur for
fruitful discussions.
The work of A.G.M. on the project
``Nuclear collective dynamics for high temperatures and neutron-proton
asymmetries'' was
supported by the Program
``Fundamental research in high energy physics and nuclear physics
(international collaboration)''
at the Department of Nuclear Physics and Energy of the National
Academy of Sciences of Ukraine.
The work of M.I.G. was supported
by the Program of Fundamental Research of the Department of
Physics and Astronomy of the National Academy of Sciences of Ukraine.


\begin{thebibliography}{99}

\bibitem{LLv6} L.D.\ Landau and E.M.\ Lifshitz, {\it Hydrodynamics,
 Course of
Theoretical Physics}, (Nauka, Moscow, 2000), Vol. 6.

\bibitem{To66}
K.B. Tolpygo, {\it Thermodynamics and statistical physics},
(Kiev University, 1966).

\bibitem{Fe72} J.H.~Fertziger and H.G.~Kaper,  {\it Mathematical Theory
of Transport Processes in Gases} (North-Holand, Amsterdam, 1972).

\bibitem{chapman} S.\ Chapman and T.G.\ Cowling,
{\it The Mathematical Theory of Non-Uniform Gases}
(Cambridge University Press. Cambridge, 1952).

\bibitem{LPv10}
E.M.\ Lifshitz and L.P.\ Pitajevski, {\it Physical Kinetics, Course of
Theoretical Physics} (Nauka, Moscow, 1981), Vol. 10.

\bibitem{LLv5} L. D.~ Landau and E. M.~ Lifshitz, {\it Statistical Physics}
  (Oxford: Pergamon) 1975.

\bibitem{Ce69} C.~Cercignani, {\it Mathematical Methods in Kinetic Theory},
  (Plenum Press, New York) 1969.

\bibitem{Ce75} C.~Cercignani, {\it Theory and application of the
  Boltzmann equation} (Scottish Academic Press, Edinburg and London), 1975.
  

\bibitem{Gr56} M. Greenspan, J.\ Acoust.\ Soc.\ Am.\ {\bf 28} (1956) 644.

\bibitem{Gr65book} M.\ Greenspan, in {\it Physics Acoustics}, edited by
W.P. Mason (Academic, New York, 1965), Vol. II.

\bibitem{schotter74} R.~Schotter, Phys. Fluids, {\bf 17} (1974) 1163.

\bibitem{Me57} M.E.~Meyer and G.~Sessler, Z.\ Phys.\  {\bf 149} (1957) 15.

 \bibitem{WaCh48} C.S.~Wang Chang, Report APL/JHU CM-467, UMH-3-F, Dept. of Engineering Research, University of Michigan  (1948).

\bibitem{WaCh52} C.S.~Wang Chang and G.E. Uhlenbeck, Project M999,
  Engineering Research Istitute,  University of Michigan (1952)
  [Reprinted in: Studies in Statistical Mechanics, edited by J. de Boer and G.E. Uhlenbeck, Vol. V, pp. 43-75, Amsterdam, 1970].

\bibitem{Wo80} L.C.\ Woods and H. Troughton, J. Fluid. Mech., {\bf 100}
(1980) 321.

\bibitem{Le89} G.\ Lebon and A. Cloot, Wave Motion, {\bf 11} (1989) 23.

\bibitem{Ko67} M.N. Kogan, {\it Dynamics of the Dilute Gas. Kinetic Theory},
  (Nauka, Fizmatlit, Moscow, 1967).

  \bibitem{BGK54}
P.L.\ Bhatnagar, E.P.\ Gross, and M.\ Krook,
Phys. Rev.,{\bf 94} (1954) 511.

\bibitem{GJ59}
  E.P.~ Gross, and E.A.~ Jackson, Phys. Fluids, {\bf 2}, 432 (1959).

  \bibitem{He93} H. Heiselberg, C.J. Pethick, and D.G. Revenhall, Ann.
Phys. (N.Y.), {\bf 223}, 37 (1993).

  \bibitem{Ta10} M.~Takamoto and S.~Inutsuka, Prog. Theor. Phys.,
  {\bf 123} (2010) 903.

  \bibitem{Ma14} A.G.\ Magner, D.V.\ Gorpinchenko, and J.\ Bartel,
Phys.\ At.\ Nucl., {\bf 77} (2014) 1229.

\bibitem{Si65} L.~Sirovich and J. K.~Thurber, J. Acoust. Soc. Am.,
  {\bf 37}, 329 (1965).

\bibitem{Bu66} J.K.~Buckner and J.H.~Ferziger, Phys. Fluids, {\bf 9},
  2315 (1966).

\bibitem{Th79} J.R.~ Thomas, Jr. and C.E.~Siewert, Thansport Theory and Statistical Physics, {\bf 8}, 219 (1979).
  
 \bibitem{Lo79} S.K.~ Loyalka and T.C.~Cheng, Phys. Fluids, {\bf 22},
   830 (1979).

 \bibitem{Ch81} T.C.~Cheng and S.K.~ Loyalka, Prog. Nucl. Energy,
   {\bf 8}, 263 (1981).

   \bibitem{Ga06} R.D.M.~Garcia and C.E.~Siewert, Z. angew. Math. Phys.
     {\bf 57}, 94 (2006.

   \bibitem{Sh08} F.~Sharipov and D. Kalempa, J. Acoust. Soc. Am., {\bf 124},
     1993 (2008).   

\bibitem{Be61} I. L.~ Bekharevich and I. M.~ Khalatnikov, Sov. J. Exp. Theor.
  Phys. {\bf 12}, 1187 (1961).

  \bibitem{Iv81} Yu. B.~ Ivanov, Zh. Eksp. Teor. Fiz., Nucl. Phys. A
    {\bf 365}, 301 (1981).

  \bibitem{Ma87} A. G.~ Magner, Sov. J. Nucl. Phys. {\bf 45}, 235 (1987).

    \bibitem{Ha01} N.G. Hadjiconstantinou and A.L. Garcia, Phys. of Fluids,
{\bf 13} (2001) 1040.

  \bibitem{St87} V.M.~Strutinsky, S.M.~Vydrug-Vlasenko, and A.G.~Magner,
    Z. Phys. A {\bf 327}, 267 (1987).

  \bibitem{Ko94} V. M.~ Kolomietz, A. G.~ Magner, V. M.~ Strutinsky, and
      S. M.~ Vydrug-Vlasenko, Nucl. Phys. A {\bf 571}, 117 (1994).


\bibitem{MGGpre17} A.G.\ Magner, M.I.\ Gorenstein, and U.V.\ Grygoriev,
Phys. Rev. E, {\bf 95} (2017) 052113.

\bibitem{MGGpre17a} A.G.\ Magner, M.I.\ Gorenstein, and U.V.\ Grygoriev,
Phys. Rev. E,  {\bf 96} (2017) 062142.

\bibitem{Fo75} D.~Forster,  {\it Hydrodynamic Fluctuations,
Broken Symmetry and Correlations Functions} (Benjamen, London, 1975).

\bibitem{Zu02} D.\ Zubarev, V.\ Morozov, G.\ R\"opke, {\it
Statistical Mechanics
of Non-Equilibrium Processes} (Fizmatlit, Moscow, 2002), Vol.~II.

\bibitem{Si71} V.P.\ Silin, {\it Introduction to the Kinetic Theory of
Gases} (Nauka, Moscow, 1971).



\end{thebibliography}
\end{document}